\input{aipcheck}

\documentclass[
    ,final            
  ]
  {aipproc}

\layoutstyle{8x11single}

\begin{document}

\title{Top Quark Physics at the LHC}


\author{J.D'Hondt}{
  address={Vrije Universiteit Brussel, Pleinlaan 2, 1050 Brussel, Belgium}
}

\begin{abstract}
The Large Hadron Collider (LHC) is expected to provide proton-proton collisions at a centre-of-mass energy of 14 TeV, yielding millions of of top quark events. The top-physics potential of the two general purpose experiments, ATLAS and CMS, is discussed according to state-of-the-art simulation of both physics and detectors. 
An overview is given of the most important results with emphasis on the expected improvements in our
understanding of physics connected to the top quark.
\end{abstract}

\maketitle


\section{Proton Collisions at the LHC}

The Large Hadron Collider or LHC is to become operational after the summer of 2007. The proton-proton accelerator will
provide collisions at 14 TeV with a luminosity starting at $2 \cdot 10^{33} cm^{-2}s^{-1}$ up to $10^{34} cm^{-2}s^{-1}$ in a later stage.
This will result in an integrated luminosity of about 10fb$^{-1}$ per year. 

Two general purpose detectors, ATLAS and CMS, will collect the final state events of from head-on proton-proton collisions taking place every 25 ns.
This defines challenging constraints on the trigger systems of the experiments. In this dense particle environment both detector designs are optimized 
with respect to the performance of for example vertex reconstruction, track momentum measurement up to high energies and
calorimetry segmentation for jet reconstruction. Therefore the LHC experiment and its detectors offer perfect settings to study the 
top quarks produced in the proton collisions.

The following sections describe the physics results in the domain of top quark physics one can potentially achieve with this experiment.
It provides both a data set to pin down the Standard Model measurements connected to this interesting fermion and, complementary, a unique window to new physics.
The increase in collision energy with respect to the Tevatron accelerator provides a natural decrease of background processes relative to
top quark signal processes.

The potential of the LHC data is been determined via simulated events reflecting the state-of-art knowledge of our detector performance.
Usually the leading-order PYTHIA generator has been used, while for some specific studies dedicated generators like AlpGen or CompHEP have been applied.
Next-to-leading order corrections have been accounted for where needed. Because the detector responds has been simulated with GEANT-4, major
data-challenge efforts have been made to produce dedicated simulation. A number of pile-up collisions are added as minimum bias events with a Poissonian 
expected value reflecting the low luminosity settings of the accelerator, namely $2 \cdot 10^{33} cm^{-2}s^{-1}$.
The information concerning the reconstruction of the general physics objects can be found on the respective webpages of the experiments:
\url{http://atlas.web.cern.ch} and \url{http://cms.cern.ch/iCMS}.

\section{Cross Section measurements}

According to the Standard Model the dominant top quark production is either in pairs or single in proton collisions. The top quark pair production has
already been observed and measured at a collision energy of 1.8-1.96 TeV at the Tevatron, while the single top quark production is yet to be
discovered. The top quark has a branching ratio of about 100\% to decay as $t \rightarrow bW$. Therefore its decay channels are defined by the decay
channels of the W boson, neglecting QCD corrections this is BR($W \rightarrow l\nu_l$)=$\frac{1}{3}$ and BR($W \rightarrow q'{\bar q}$)=$\frac{2}{3}$.

\subsection{Top Quark Pair production}

The next-to-leading order cross section for top quark pair production in proton collisions at 14 TeV is calculated to be 833pb 
of which some references are given in \cite{nason}. This would result in more than 8 million $pp \rightarrow t{\bar t}$ events per year in the low luminosity
settings of the experiment or about 1 such event per second to be compared with about 10 per day at the Tevatron.
Contrary to the Tevatron proton-antiproton collisions at 1.96 TeV at the LHC the production mechanism 
is with 87\% dominated by gluon fusion.

The selection of the events in the three categories of $t{\bar t}$ decay channels is based on the reconstruction and calibration of jets, b-tagging algorithms and
for the leptonic decays on the isolation of the lepton. Each channel has different trigger requirements and different dominating backgrounds. Applying both
kinematic cuts on the reconstructed objects and topological cuts, the following efficiencies $\epsilon$ and signal-to-noise ratios (S/N) are reaches:

\begin{itemize}
\item Fully hadronic channel ($t{\bar t} \rightarrow bq'{\bar q}{\bar b}Q'{\bar Q}$, 3.7M events per 10 fb$^{-1}$): the dominating background 
are the QCD multi-jet events from general $2 \rightarrow 2$ processes with radiation, for a S/N ratio of $\frac{1}{9}$ and value $\epsilon \simeq$ 2.7\% was reached from the
fast simulation of ATLAS \cite{ATLAStopmass}.
\item Semi-leptonic channel ($t{\bar t} \rightarrow bq'{\bar q}{\bar b}l\nu_l$, 2.5M events (l=e,$\mu$) per 10 fb$^{-1}$): the dominating background are the 
W+$n$jet events, for a S/N ratio of about 27 a value of $\epsilon \simeq$ 6.3\% was reached from the full simulation of CMS \cite{CMSxspair,CMStopmass}.
\item Fully leptonic channel ($t{\bar t} \rightarrow bl'\nu_{l'}{\bar b}l\nu_l$, 0.4M events (l=e,$\mu$) per 10 fb$^{-1}$): the Drell-Yan process and the Z+$n$jet events
are the dominating backgrounds which can be reduced to a S/N ratio of about 5 for an efficiency of $\epsilon \simeq$ 5\% \cite{ATLAStopmass,CMSpairother}.
\end{itemize}

From the selected signal and background events in the simulation an estimate was made of the precision which can be obtained on the cross section
with an integrated luminosity of 10 fb$^{-1}$. Several 
systematic uncertainties are taken into account, amongst other they are related to the pile-up, the jet energy scale calibration, the radiation effects,
the b-tagging (in)efficiencies, the parton density functions and the uncertainty on the integrated luminosity \cite{CMSxspair,CMSpairother}:

\begin{itemize}
\item Fully hadronic: $\frac{\Delta \sigma}{\sigma}$ = 3 (stat) $\pm$ 18 (syst) $\pm$ 3 (lumi) \% 
\item Semi leptonic (e/$\mu$): $\frac{\Delta \sigma}{\sigma}$ = 0.4 (stat) $\pm$ 9.2 (syst) $\pm$ 3 (lumi) \% 
\item Fully leptonic (e/$\mu$): $\frac{\Delta \sigma}{\sigma}$ = 0.9 (stat) $\pm$ 11 (syst) $\pm$ 3 (lumi) \% 
\item Fully leptonic ($\tau$): $\frac{\Delta \sigma}{\sigma}$ = 1.3 (stat) $\pm$ 16 (syst) $\pm$ 3 (lumi) \% 
\end{itemize}

The dominating uncertainty is related to the knowledge of the b-tagging performance which can be extracted from the $t{\bar t}$ data itself, as shown
in a novel CMS method \cite{CMSbtagtt}. These total uncertainties on the $t{\bar t}$ cross section reveal that an indirect top quark mass measurement is possible via
the cross section to a precision of about 2-3 GeV/c$^2$, which is already in the same precision range compared to the current Tevatron combined measurement.
It was shown in \cite{ATLAStopmass,CMSxspair} that shape analyses like those performed at the Tevatron \cite{D0shape} which combine topological information
of several observables, are not directly useful at the LHC for the determination of the $t{\bar t}$ production cross section. They can however be applied in 
detailed studies of the radiation processes when measuring exclusive cross sections. 

\subsection{Single Top Quark production}

The total next-to-leading order cross section for single top and anti-top production is 373 pb at the LHC of which some references are given in \cite{stelzer}.
In this total cross section the fraction from top and anti-top is not equal. The events can be produced in three channels:

\begin{itemize}
\item s-channel: $\sigma^{NLO}(t)$ = 6.6 pb and $\sigma^{NLO}({\bar t})$ = 4.1 pb;
\item t-channel: $\sigma^{NLO}(t)$ = 153 pb and $\sigma^{NLO}({\bar t})$ = 90 pb;
\item associated tW-channel: $\sigma^{NLO}(t)$ = 60 pb and $\sigma^{NLO}({\bar t})$ = 60 pb.
\end{itemize}

This is a significant increase compared to the cross section of the same processes at the Tevatron, while the backgrounds like $Wjj$ do not increase much
from Tevatron to LHC settings. Each channel is sensitive to different signals of new physics: the s-channel to heavy W' bosons, the t-channel to Flavour-Changing
Neutral Currents (FCNC) and the tW-channel to charged Higgs bosons.

The event selections depend on the channel, but are based on missing transverse momentum, b-tagging and topological cuts reflecting the expected kinematical topology
of the reconstructed event. In many cases control regions for the background level estimation are developed reducing the overall systematic uncertainty due to the
uncertainty on the background cross section. The following potential results are obtained in the cross sections \cite{CMSsingletop,ATLASsingletop}:

\begin{itemize}
\item t-channel: $\frac{\Delta \sigma}{\sigma} = 2.7 (stat) \pm 8.1 (syst) \pm 3 (lumi) \%$ (CMS for 10 fb$^{-1}$);
\item s-channel: $\frac{\Delta \sigma}{\sigma} = 18 (stat) \pm 31 (syst) \pm 3 (lumi) \%$ (CMS for 10 fb$^{-1}$);
\item s-channel: $\frac{\Delta \sigma}{\sigma} = 12 (stat) \pm 10.1 (syst.exp) \pm 8 (bck. theo) \pm 5 (lumi) \%$ (ATLAS for 30 fb$^{-1}$).
\end{itemize}

For the tW-channel CMS obtains in the di-lepton and single-lepton decay channel a combined sensitivity of 6.4$\sigma$ at 10 fb$^{-1}$
for observing the process. The limiting factor
being the systematic uncertainties due to the jet energy scale and the b-tagging knowledge, both can be improved when a better understanding of the true
detector performance is reached.

The measurement of the cross sections are directly related to the $|V_{tb}|$ matrix element in the CKM matrix. In this measurement the s-channel is preferred because
the t-channel is dominated by uncertainties from the parton density functions.

\section{Top Quark mass}

\begin{figure}
  \includegraphics[height=.315\textheight]{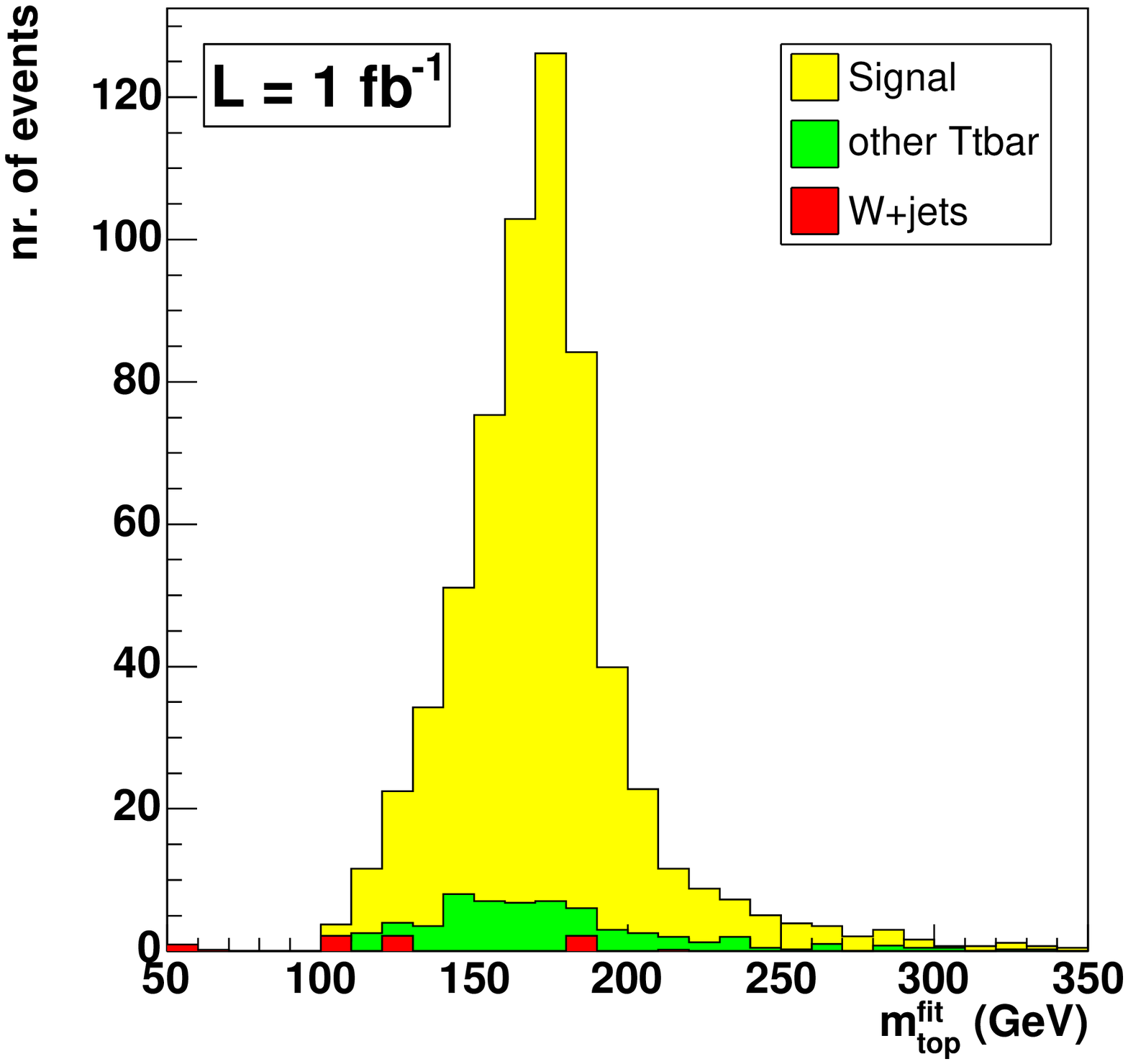}
  \includegraphics[height=.3\textheight]{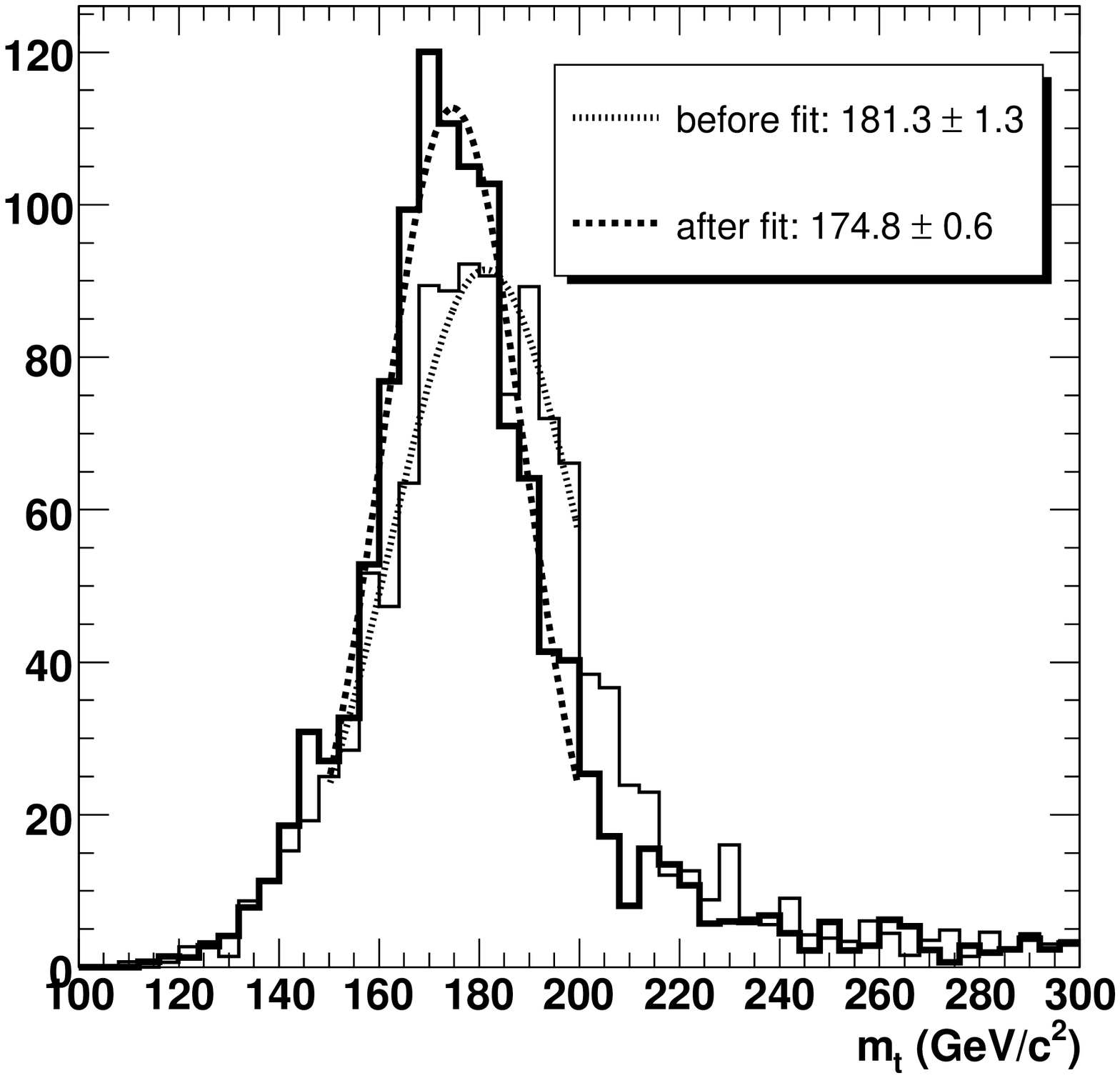}
  \caption{Left: Distribution of the mass of the hadronic decaying top quark reconstructed in the CMS detector for the selected events after applying the kinematic fit.
The contribution of all relevant background processes is shown. Right: Comparison of the reconstructed and fitted top quark mass distribution from CMS using only
correct jet combinations.}
\label{fig:kinfit}
\end{figure}

One of the most important measurements to be performed at the LHC amongst all Standard Model measurements \cite{SMproc} 
is a precision measurement of the mass of the top quark. This will happen from the top quark pair
events in which after the event selection several challenges remain: the different jet combinations, the influence of the background and the missing neutrino's.
The selection of the correct jet pairing can be enhanced by examining on an event-by-event basis the kinematic topology of the selected event. Several
topological observables which characterize a jet combination can be combined \cite{CMStopmass}, for example the angle between the b jet in the 3-jet system reflecting the
top quark and the lepton \cite{ATLAStopmass}. Kinematic fits can be applied forcing the W boson mass to the 2-jet system from its decay \cite{CMSkinfit}. This reduces 
the influence of the jet energy scale calibration \cite{ATLAStopmass,CMStopmass} and increases the statistical power of the event, see Figure~\ref{fig:kinfit}.

In \cite{CMStopmass} advanced statistical inference techniques are applied to extract the top quark mass from the selected events. Following the LEP and Tevatron
expertise \cite{TeVideo, DELPHI} convolution methods are introduced which weight the reconstructed ideogram reflecting all kinematic 
information in the event with the expected theoretical
probability density function which are basically Breit-Wigner functions. This results in a likelihood per event which can be weighted according to the observed
topology in the event enhancing for example signal rather than background events or correct jet combinations rather than wrong ones.

It is shown in \cite{CMStopmass} that in the so-called golden channel ($t{\bar t} \rightarrow bq'{\bar q}{\bar b}\mu\nu_{\mu}$) a total uncertainty
of 1 GeV/c$^2$ could be reached with 10 fb$^{-1}$ of data. The uncertainty is totally dominated by systematic uncertainties due to the jet energy 
scale calibration. To reach the benchmark precision of 1 GeV/c$^2$ on the top quark mass, a 1.5\% uncertainy is to be reached on the b jet energy scale
calibration. The light quark jets can be calibrated directly from the W boson mass constraint in $t{\bar t}$ events as shown in \cite{CMScaljet}.

A promising alternative for a direct measurement of the top quark mass is introduced in \cite{JPsiintro} and explored with either full or fast detector
simulation in \cite{ATLAStopmass,CMSJPsi}. From the semi-leptonic $t{\bar t}$ decay followed with the decay $t \rightarrow l + J/\psi + X$ (where $J/\psi \rightarrow \mu\mu$)
the invariant mass of the 3-lepton system is due to the large mass of the $J/\psi$ correlated with the top quark mass. Although one can extract a very clean
sample of this signal, due to the small branching ratios involved this study has to be performed with at least 10 fb$^{-1}$ of data. The expected statistical uncertainty
on the top quark mass drops below 1 GeV/c$^2$ from an integrated luminosity of 20 fb$^{-1}$ onwards. This method is in first order not polluted from uncertainties related
to the definition of the reconstructed jets. Most of the systematic uncertainties are however related to the theoretical modeling of the events. They add up to about 
1.5 GeV/c$^2$, triggering that more work is needed to understand these phenomena.

Work is also needed to determine the correlation between the divers range of methods and the different channels explored at the LHC to extract the top quark mass. 
With a proper understanding of the detector performance a precision on the top quark mass of 1 GeV/c$^2$ is however feasible.

\section{Spin correlations}

The top quark does not loose its spin information before it is decaying. Therefore measurements can be performed to understand the spin correlations between
both top quarks in the pair produced events. They have been implemented in TopRex for example \cite{toprex}.
The measurement of possible asymmetries ${\cal A}$ is performed from the double differential distribution of the decay angle $\theta$:

\begin{equation}
\frac{1}{N} \frac{d^2N}{d cos\theta_l d cos\theta_q} = \frac{1}{4} ( 1 - {\cal A} \kappa_l \kappa_q cos\theta_l cos\theta_q) 
\end{equation}

\noindent
where the two observables are the angle $\theta_l$ ($\theta_q$) between the top (anti-top) direction in the $t{\bar t}$ center-of-mass frame and the lepton (quark)
direction of flight in the top (anti-top) rest frame. Several combinations of spin analyzers are studies, but the spin analyzing power $\kappa$ is
strongest for the lepton and least energetic quark couple \cite{thspin}: ${\cal A} \kappa_l \kappa_q$ = 0.158 according to the Standard Model prediction
in next-to-leading order. 

The potential for spin correlations measurements is studied both in ATLAS \cite{ATLASspin} and CMS \cite{CMSspin}.
For 10 fb$^{-1}$ CMS developed an measurement which results in:

\begin{itemize}
\item ${\cal A}_{b-l} = 0.375 \pm 0.027 (stat)^{+0.055}_{-0.084} (syst)$
\item ${\cal A}_{q-l} = 0.346 \pm 0.021 (stat)^{+0.026}_{-0.055} (syst)$
\end{itemize}

\noindent
with central values in good agreement with the prediction as implemented in TopRex. In ATLAS a analysis has been performed using single differential
distributions of the opening angle between two spin analyzers. For 10 fb$^{-1}$ a sensitivity of around 5$\sigma$ was obtained to observe the correlations 
according to the Standard Model predictions. The main systematic uncertainties are related to the parton generation (parton density functions and the
$Q^2$ scale dependency), the radiation and the jet energy scale uncertainties.

\section{W helicity in Top Quark decays}

The large top quark mass allows the W boson in the top quark decay to be longitudinaly polarized. This could be confirmed by 
a measurement of the angular distribution $\theta$ which is the angle
between the lepton in the W boson rest frame and the W boson in the top quark rest frame.
The differential distribution is defined as:

\begin{equation}
\frac{1}{N} \frac{dN(W \rightarrow l\nu_l)}{d cos\theta} = K \left[ f_0 m_t^2 sin^2\theta + f_L m_W^2 (1-cos\theta)^2 + f_R m_W^2 (1+cos\theta)^2 \right] 
\end{equation}

\noindent
where the Standard Model prediction of $f_0 = \frac{m_t^2}{2m_W^2 + m_t^2} \simeq 0.7$ and $f_R \simeq 0$ (when $m_b \simeq 0$).
The potential for measuring $f_L$, $f_0$ and $f_R$ has been determined by ATLAS \cite{ATLASspin} for a dataset of 10 fb$^{-1}$. They are transformed
into measurements of angular asymmetries, $A_{FB} = \frac{3}{4} (f_L - f_R)$, $A_+ = -3\beta (f_0 + (1+\beta) f_R)$ and $A_- = 3\beta (f_0 + (1+\beta) f_L)$
with $\beta = 2^{\frac{2}{3}}-1$. The results are deconvoluted for detector effects via simulation:

\begin{itemize}
\item $A_{FB} = 0.2234 \pm 0.0035 (stat) \pm 0.0130 (syst)$
\item $A_+ = -0.5472 \pm 0.0032 (stat) \pm 0.0099 (syst)$
\item $A_- = 0.8387 \pm 0.0018 (stat) \pm 0.0028 (syst)$
\end{itemize}

\noindent
where the systematic uncertainties dominate the statistical precision. Potential confidence intervals at the 2$\sigma$ level are obtained on the
general Wtb vertex which reach significant better limits compared to the Tevatron: $-0.14 < V_R < 0.19$, $-0.10 < g_L < 0.07$ and $-0.04 < g_R < 0.04$.

\section{Searches for New Physics}

In many directions one can search for deviations of the Standard Model predictions in the selected top quark events. The appearance of FCNC phenomena is suppressed in the 
Standard Model with a factor 10$^{-12}$, while in extensions of the Standard Model (like in supersymmetric models) they could be less suppressed to the level
of only 10$^{-4}$. Three couplings are studied in \cite{ATLASFCNC}: $t \rightarrow qZ$, $t \rightarrow q\gamma$ and $t \rightarrow qg$. For each of the channels a
topological likelihood variable is constructed to identify the signal above the Standard Model background. The results are interpreted as potential limits on the
branching ratios of the decays. Already at 10 fb$^{-1}$ they reach a level which is two orders of magnitude better (lower) than the current Tevatron 
extrapolated predictions for 2 fb$^{-1}$.

Also in the domain of flavour physics the potential of a search for like-sign top quark pairs and other FCNC signals have been explored by CMS \cite{CMSsign}.
From top- and technipions in topcolor assisted Technicolor models or gluino pair production in the MSSM, a signal could be created which results in top quark pairs
with the same electric charge. Hence in the di-lepton decay channel this would be visible by measuring the charge of the reconstructed leptons. When using a 
ratio R of selected events with the same versus different lepton charges, most of the systematic uncertainties cancel and the new physics models can be probed.
With an integrated luminosity of 30 fb$^{-1}$ a cross section of like-sign $pp \rightarrow tt ({\bar t}{\bar t})$ events of 1 pb becomes visible as a 5$\sigma$ effect
on this ratio R.

The reconstructed invariant mass distribution of the $t{\bar t}$ system is an important window to new physics. Technicolor and supersymmetry based models
can introduce a resonance in this distribution. For example MSSM one-loop electro-weak corrections give a significant distortion in the shape
of this distribution. For heavy Higgs $A^0$ masses from the $t{\bar t}$ production threshold of about 350 GeV/c$^2$ up to about 500 GeV/c$^2$ this
distortion could be visible as it is much larger than the Standard Model corrections \cite{LHCyellow}.

\section{Commissioning with Top Quarks}

Top quark events at the LHC are very complex and their reconstruction is based on almost each crucial aspect of the detector. Therefore
top quark events are already from the start-up of the experiment perfect for the commissioning of the detector and the physics performance of the new detectors.
The measurement of the b-tagging efficiency \cite{CMSbtagtt} and the measurement of the jet energy scale calibration factors \cite{CMScaljet} 
can be performed on the events from the first weeks of data taking. It is shown that with an integrated luminosity of about 300 pb$^{-1}$ the jet energy scale
can be extracted using the W boson mass constraint in the event to a precision of better than 1\%. The main challenge being the control of the pile-up 
collisions, for which vertex counting methods can be envisaged.

In general, a rediscovery of the top quark peak at the LHC would be one of the first milestones in the physics commissioning of the detector performance \cite{ATLAScomm}.
The top quark pair production cross section will be one of the first relevant measurements to be performed. It is shown that in a few hours of data taking and without 
b-tagging the W boson and the top quark mass peaks are visible above the background. With the first day of running (30pb$^{-1}$) the top quark mass could be
measured to a precision of 3 GeV/c$^2$ and the W boson mass to 1 GeV/c$^2$. This will allow for a first estimation of the jet energy scale calibration corrections and
a first performance test of the b-tagging algorithms. The main challenge is to select a pure top quark sample without the application of b-tagging identification
algorithms.

\section{Conclusion and Outlook}

It is demonstrated with the use of state-of-art physics and detector simulation that the LHC proton collision data opens 
a new area of top quark physics. Already from the data collected in the first weeks of the LHC experiment the top quark measurements
are of interest in the physics commissioning of the individual experiments.
With the huge amount of statistics the production mechanisms of top quark processes, 
the particles properties and its decay can be studied for the first time in great detail. For most of the measurement
analyses the challenge will be in the domain of the systematic uncertainties related to the detector and the soft-QCD involved in the
collisions. For the analyses searching for deviations from the Standard Model predictions in the above one becomes sensitive
to a wide range of physics models beyond the Standard Model, an area to be further explored in the high luminosity settings of the LHC experiment.


\begin{theacknowledgments}
 The author has presented this overview on behalf of the LHC Collaborations and wishes to thanks those who have
contributed to this excellent research.
\end{theacknowledgments}



\bibliographystyle{aipproc}   

\bibliography{sample}


\end{document}